Using Information Governance to Evaluate Patient Care in Amazon's One Medical

Andrew M. Nguyen, MS, NREMT

Department of Public Health and Community Medicine, Tufts University School of Medicine, Boston, MA, USA

**1.1 Abstract**

This paper explores the information governance (IG) maturity of Amazon's One Medical (AOM), a digital health and telehealth primary care organization. Combining Amazon's technology expertise with One Medical's healthcare services, AOM aims to transform healthcare through a human-centered, technology-powered model. However, successfully integrating Amazon's disruptive approach into the complex healthcare industry is challenging. To examine AOM's IG maturity, we compare and critique patient care in AOM against the ARMA Maturity Model and AHIMA's IGPHC framework. The study discusses the implications of telehealth on the doctor-patient relationship, rising roles of ancillary service teams, commoditization of healthcare, and potential monopolization. It also addresses security risks, compliance challenges, and the impact of technology on disadvantaged populations. Our analysis highlights the growing importance of information management in the evolving intersection between healthcare and technology and suggests potential areas for improvement in AOM's IG maturity.

**2.1 Introduction**

One Medical is a digital health and telehealth primary care organization that combines in-person care with virtual care services. Its mission is "to transform health care for all through a human-centered, technology-powered model" (Amazon and One Medical sign an agreement for Amazon to acquire One Medical, 2022). One Medical parent 1Life Healthcare (NASDAQ: ONEM) operates more than 180 medical offices across the US in collaboration with over 8000 companies (Evans & Herrera, 2022). On July 21, 2022, 1Life Healthcare announced a $3.9 billion merger with e-commerce giant Amazon (NASDAQ: AMZN) under which Amazon will acquire One Medical. Neil Lindsay, Senior Vice President of Amazon Health Services, commented that, together, the two organizations will "help more people get better care, when and how they need it" (Amazon and One Medical sign an agreement for Amazon to acquire One Medical, 2022).

Across the world, people looked to Amazon's bold social and cultural mythos of success to imagine its future in healthcare. Chamath Palihapitiya, Canadian venture capitalist, quipped that patients may soon become enmeshed in Amazon's loop of services: "Your One Medical doctor will provision a blood test. He or she will analyze it over telehealth. They will prescribe a better diet. That will be sent to Whole Foods" (Chamath et al., 2022, 1:36:36). On Twitter, users jumped to envision how the acquisition may transform the healthcare market with Amazon's ethos on simplicity and ease-of-use (Exhibit 1) (Mortensen, 2022).

However, the integrated virtual primary care vision may be the technology leader's most challenging expansion yet (Evans & Herrera, 2022). Most technology companies that venture into healthcare ultimately fail. Indeed, bringing Amazon's disruptive convenience, ambitious experimentation, and synergistic business models into one of the most traditional, complex, and difficult industries is an ambitious goal (Huckman & Staats, 2022). Making healthcare work in tandem with Amazon's growing e-commerce, cloud computing, digital streaming, and artificial intelligence brands is an even greater one.

Amazon's future healthcare offering, herein described as Amazon's One Medical (AOM), is still nascent and has yet to fully develop. As a leading technology and e-commerce company, it may be well positioned to manage and govern information and data across businesses and industries. However, information and data governance from a healthcare perspective may offer further insights and pathways to success. How does it compare to current standards and policies? In particular, how mature will AOM's information governance be? Will its governance help or hinder its mission? How can it improve? To elucidate these gaps, patient care in AOM will be compared and critiqued against the ARMA Maturity Model and AHIMA's IGPHC.

**2.2 Information Governance**

Information governance (IG) is defined as: "an organization-wide framework for managing information throughout its lifecycle and supporting the organization's strategy, operations, regulatory, legal, risk, and environmental requirements" (Empel, 2014). It is a set of best practices that guide an organization's information management strategy in wide-ranging, long-term contexts that are strategic, tactical, and proactive. Review groups made up of medicine, nursing, information technology, law, privacy, security, and financial professionals work together to create these IG principles. In the context of increasing volume, variety, veracity, velocity, and value (5 V's) in healthcare, IG provides powerful tools to consider the needs of all healthcare stakeholders, improving organization-wide success and reducing systemic risk. (Empel, 2014).

In the context of patient care, IG may improve care quality, outcomes, safety, efficiency, efficacy, and costs. Specifically, a review on healthcare case studies concluded that organizations employing IG programs improved the quality of patient outcome tracking, increased patient data accuracy, boosted patient engagement through data sharing, invited more collaboration between providers during the documentation process, promoted predictive analytics, and decreased costs by removing redundant workflows. However, IG was observed as an enabler to success on top of executive leadership and support (Lesley, 2014).

AOM introduces patients as another rising core stakeholder in patient care. Examining the relationship that patients share with not only their own health, but with the IG of disruptive healthcare and technology innovation may also further enable improvements in care.

**2.3 ARMA Maturity Model**

ARMA International is an organization that promotes and advocates for information management guidelines for professionals. Its principles outline best practices to help organizations learn, grow, and succeed. The organization's Principles Maturity Model is based on Generally Accepted Recordkeeping Principles (GARP) that can be used as a quality improvement tool (The Principles Maturity Model, 2022). The maturity model outlines five levels of governance maturity: level 1 (sub-standard), level 2 (in development), level 3 (essential), level 4 (proactive), and level 5 (transformational). The five stratifications demonstrate how well duties, values, and information assets intersect between stakeholders in an organization (How the Information Governance Reference Model (IGRM) Complements ARMA International's Generally Accepted Recordkeeping Principles (GARP®), 2011).

**2.4 AHIMA's IGPHC**

Similar to ARMA International, American Health Information Management Association (AHIMA) is a global nonprofit organization that advocates for trusted health information between people, systems, and ideas (About Us, 2022). AHIMA's Information Governance Principles for Healthcare (IGPHC) is an IG framework specific to healthcare that outlines eight key principles: accountability, transparency, integrity, protection, compliance, availability, retention, and disposition (Empel, 2014). Each key principle may be analyzed in the context of the ARMA Maturity Model, providing in-depth IG metrics unique to each principle. This synergy provides organizations with a check list of strengths, weaknesses, and opportunities for improvement (How the Information Governance Reference Model (IGRM) Complements ARMA International's Generally Accepted Recordkeeping Principles (GARP®), 2011).

## 3.1 Methods

To evaluate AOM in the context of the ARMA Maturity Model and AHIMA's IGPHC, we propose three hypothetical scenarios in which patients encounter, interact, and navigate its future telehealth and health services. AOM services were anticipated based on a reflection of the sociocultural mythos of Amazon by Chamath, Mortensen, and Huckman & Staats (2022; 2022; 2022). Though many stakeholders exist in primary care, we opt to align with Amazon's vision to "focus relentlessly" on customers and highlight the experiences of healthcare's most important stakeholder: the patient (Barr, 2021). Personas of each patient were ideated based on core demographics of patients who use primary care services. Scenarios of each persona were created in the context of a trigger event that compelled them to use AOM services. Their journeys may be positively, negatively, or not at all affected by IG. Each of the eight key principles from AHIMA's IGPHC will draw from the scenarios to assess AOM in terms of the ARMA Maturity Model.

## 4.1 Results

Prior to ideation, a literature review was performed to survey who visits primary care practices. The Robert Graham Foundation found that patients from all ages with acute and chronic conditions seek primary care (Petterson et al., 2018). However, most patients tend to either be younger patients (0-4 years) or older patients (65-100). Both groups average around 2.5 to 3 visits a year, and females had more visits per year (1.65) than that of males (1.44). Furthermore, more patients with either hypertension, diabetes, and asthma, or a combination of other chronic diseases visited primary care physicians than patients with other chronic conditions

(Agency for Healthcare Research and Quality, 2014). In terms of health insurance, younger patients are more likely to participate in Medicaid, while older adults are more likely to use Medicare. However, over half of all primary care patients have commercial insurance (Medical Group Management Association (MGMA), 2015). With demographics in mind, three personas, triggers, and scenarios were formulated.

**4.2 Persona 1: Ellis**

Ellis, a military veteran who is now a retired carpenter, lives with his wife in Austin (Table 1). He has hypertension and diabetes. In his spare time, he coaches a soccer team for kids.

During practice, Ellis suffers a heart attack. He promises his wife that he will be more accountable for his medications and diet regimen. He looks to AOM for a convenient, virtual solution to stay healthy.

As an Amazon Prime member, Ellis searches for primary care providers in his area on Amazon.com. The marketplace shows the name, location, price range, ratings, and reviews for each provider. He finds that picking and choosing the highest rated and most affordable provider is much easier than sticking with the same one from his last visit. After checking out, he visits one of AOM's brick-and-mortar primary care offices for a blood test. He stays updated on prescriptions, diets, and exercises over telehealth.

One of his quotes is: "It's so easy to switch providers. I feel like I can shop for who best fits my needs on a trusted marketplace, but I do wish I had a better personal relationship with my providers."

**4.3 Persona 2: James and Angela**

The second persona reflects six-year-old James and his mother, Angela (Table 2). They are Filipino-American immigrants. To support the family, Angela is a work-from-home virtual assistant. At night, she takes classes to become a nurse.

One day, James's school asks for his vaccine and immunization records. However, most of his medical records are back in the Philippines—and they're paper, not electronic. To accommodate her busy schedule, Angela navigates AOM to arrange vaccine and immunization services for her son.

Recalling how she booked a health appointment on AOM for a client, Angela looks to the digital marketplace to get son's vaccine and immunizations updated for the coming school year. She enjoys having her son's documents on the cloud, but finds it difficult to retrieve her information from outside the US. However, she becomes concerned about AOM's convenience when realizes how many parties have access to her data across Amazon's suite of services.

Her quote is: "As a busy single-mother and immigrant, I want privacy, security, and safety for my family and me."

**4.4 Persona 3: Beverly**

Finally, the last persona is 28-year-old Beverly, a techie from San Francisco, CA who was recently let go from her job (Table 3). Her parents are unable to support her, so she couch surfs between friends nearby while searching for employment. The stress aggravates her asthma.

Without insurance, she looks to AOM for affordable primary care to work on her respiratory health. She hopes that the marketplace can match her with an empathetic provider that has time to listen to her needs.

Beverly is fortunate to have Wifi at coffee shops and her friend's apartment, but she worries that she may not have consistent internet for telehealth like she expected. Time-sensitive messages with providers are often sent or received late, which makes it difficult to coordinate care. When she finds out there is incorrect information on her record, she contacts customer service to alter it.

Her quote is: "Technology is great, but only if I am able to use it and have access to it in the first place."

## 5.1 Principle of Accountability

In scenario 1, Ellis seeks continuity of care from an accountable member of the healthcare team. He gravitates towards one point-of-contact, who, in traditional healthcare models, is the physician. However, the availability, choice, and freedom to work with different providers without barriers is too convenient to ignore. In an instant, Ellis can survey hundreds of providers alongside their price ranges, ratings, and reviews. If he encounters a negative experience, he can simply move on to another provider without hesitation. In essence, AOM creates a commoditization of providers that can be freely exchanged between patients, and vice versa. Providers are no longer accountable to patients, but to themselves—their reviews, their ratings, and their popularity.

The displacement of accountability from providers can instead be fulfilled by the Amazon customer service (CS) team. As liaisons between customers and AOM, CS representatives operate across, between, and within Amazon's suite of services, whereas providers only work at the clinic. CS teams, then, replace providers as the conduit for relationships, communication, and continuity of care. Though this paradigm departs from accountability models outside of AOM, it may relieve clinicians from current issues such as overwork and burnout, which have been demonstrated to negatively impact patient care, professionalism, and safety (West et al., 2016). Though button-up accountability from CS teams may be unconventional for most healthcare settings, it may resolve current issues that practitioners face, especially in the context of emerging technologies. As a result, Ellis loses out on relationships with his providers in exchange for ones with CS.

Amazon's reputation for strong CS teams is built on top of the world's leading cloud platform, Amazon Web Services (AWS) (What is AWS, 2022). In fact, AWS makes up 16% of Amazon's revenue and grew 37% year-over-year for Q1 2022 (Fildes, 2022). These expansive offerings are led from the top-down by senior executives such as Adam Selipsky, CEO of AWS, who have bought-in to the importance of information management. Accountability aligns with Amazon's business model to "make bold investment decisions in light of long-term leadership considerations rather than short-term profitability considerations," which will likely extend to AOM's telehealth service as well (Corporate governance, 2022). Therefore, we rate AOM with a Level 5 Transformation Maturity for the GARP Principle of Accountability.

**5.2 Principle of Transparency**

In *The Long Fix*, healthcare CEO Dr. Vivian Lee describes a transparency problem in healthcare. Patients cannot comparison shop well. Hospitals may list their prices, but their offerings are filled with complex jargon. Quality, outcomes, satisfaction, convenience, and amenities are unknown. Even providers often do not know. This opaque model disrupts competition, frustrates patients, and hurts care. Lee compares patients to "shoppers" that deserve to make informed decisions about their care (2021).

Enter AOM. Using the Amazon marketplace, Ellis can confidently "shop" for providers, care centers, and care plans using ratings, reviews, and service descriptions. He knows which ones to avoid, which ones cater best to seniors, and which ones have Spanish-speaking providers—all in the comfort of his own home. If something is missing, he can provide his own review and let other shoppers know. Or, he can reach out to the CS team to get his questions answered shortly. In this way, shoppers look out for each other and keep providers in check in an open and verifiable manner. The marketplace enables a form of game theory between patients and providers that rewards excellence and discourages anything less. If care goes well, patients and providers both win with positive experiences. Open and peer-verifiable platforms are foundational to these exchanges. When patients like Ellis derive trust from them, it contributes to a network effect of reliability that draws in more patients.

Vertical integration across Amazon's suite of services is key to transparency because it unites them under the same leadership that can design, implement, and monitor transparent systems in the first place. For example, instead of multiple invoices from different clinics with different descriptions, pricing models, and insurance plans, Ellis may receive only one invoice at AOM. It could include his blood tests, telehealth visits, and hypertension medications from one unified system. Trained staff may perform internal controls and compile statistics across these services to increase patient satisfaction, revenue, and productivity. Integration also facilitates external audits. Requests from courts and legal parties are likely to be met and fulfilled. As a result, Amazon's vertical acquisition strategies make it uniquely positioned to exhibit Level 5 Transformation Maturity for the GARP Principle of Transparency.

**5.3 Principle of Integrity**

When Ellis surveys telehealth providers on the Amazon marketplace, he notices that some are sponsored, appearing at the top of his search results. Although they may not be rated as highly, he assumes that they are authentic and reliable. After all, he trusts Amazon. In this way, partners and third parties may "buy" their way to visibility and trust, regardless of their performance metrics or history. Because sponsored services are shown alongside relevant results, Ellis may not even notice the difference.

Advertising services account for 7% of Amazon's total revenue (Fildes, 2022). As custodians of Amazon's marketplace, Amazon profits by manipulating search results with ads. The platform is not a neutral, free market, but one operated by a for-profit company that is responsible to shareholders. In other words, Amazon owns the essential infrastructure that all businesses on the platform depend on. With this monopolization, Amazon may highlight its own offerings over that of its competitors, artificially pricing them out of the market. For example, Amazon sold its Kindle device and e-books at losses to outcompete competitors, monopolizing 90% of all e-books sales by 2009 (Khan et al., 2017). Therefore, anti-competitive behavior may hurt innovation and hinder the ability for shoppers to make informed decisions.

Today, Amazon's dominance and its impact on manufacturers, consumers, and perhaps, soon—patients—is highly controversial. Although Amazon's marketplace manipulation, monopolization, and anti-competitive behaviors are its strengths, reducing them may improve patient trust. The balance between growth and reputation should be assessed frequently to prevent downstream harm to all stakeholders on the platform. As a result, we assign AOM with a rating of Level 3 Essential Maturity for the GARP Principle of Integrity.

**5.4 Principle of Protection**

Angela takes her son, James, to an AOM clinic for his vaccine and immunizations for the school year. On the way, she verifies his health record from the Amazon app. She notices some mistakes and requests to append his record through Amazon CS. As soon as she walks in, his providers view the requests and make updates alongside the vaccinations and immunizations. They finish in minutes, and follow-up through telehealth in a few days. For a different visit,

Angela takes her son to another AOM clinic nearby. His new clinicians stay up-to-date using the unified electronic health record through Amazon. Continuity-of-care is maintained. Empowered, Angela and James have many opportunities to view and interact with their health record, preventing data loss, corruption, or spoliation. Together, the family enjoys the convenience of a seamless, interoperable care experience that encourages engagement with their data.

      At the backend, AOM natively operates on top of AWS, one of the most secure cloud and data platforms available (Data Protection, 2022). However, Angela becomes concerned when she realizes how many people and organizations have access to her son's data. Beyond the point-of-care, her son's data flows to Amazon's physical stores, online stores, third-party sellers and services, advertising services, and, of course, AWS. In fact, it will be shared with a workforce that makes up 1 in 150 people in the US, or 950,000 workers as of 2021 (Reuter).

      Today, Amazon faces scrutiny over multiple data breaches and data loss events. A breach on its streaming service, Twitch, for example, leaked 125 GB of data that sparked a controversy about streamer compensation (Reuters, 2021). Preventing data breaches and maintaining confidentiality across so many complex and moving parts will become even harder as Amazon expands into healthcare, an industry that takes data security and protection seriously. Considering Amazon's sheer size, it would need the strongest IG program yet to encompass physical, technical, and non-technical safeguards.

      Yet, healthcare introduces another risky stakeholder: the patient. Angela and James' scenario demonstrates how patients may impact data protection by verifying their record. However, the responsibility to protect it also comes with risks for breach, corruption, and loss. Implementing safeguards for patients is a missing component that AOM will need to address

moving forward. With many challenges and opportunities ahead, we rate AOM with a Level 3 Essential Maturity for the GARP Principle of Protection.

## 5.5 Principle of Compliance

Foundational to Angela and James' continuous patient experience is AOM's compliance with laws, regulations, and standards across products. AWS' compliance arm, AWS Risk and Compliance, is a leader in compliance and offers compliance services. In 2022, it issued a white paper to help clients adhere to regulatory agencies locally, nationally, and internationally (Cloud Compliance). The document outlines how governance frameworks can comply with healthcare, financial services, education, government, and energy industries. Their documentation on how clients may adhere to HIPAA / HITECH regulations is especially comprehensive. It ranges from encryption, protection, auditing, backups, and disaster recovery for workloads containing patient health information (PHI) (Amazon Web Services, Inc., 2021).

Extensive compliance documentation and services underscore that compliance is at the forefront at Amazon, and will no doubt continue to play an important role at AOM. However, Amazon's entrance to healthcare is still nascent. For example, AWS documentation on HL7 FHIR, a growing interoperability framework and API, is absent (Saripalle et al., 2017). AOM should continue investing resources into compliance by creating a roadmap alongside providers, regulators, and policymakers. Communicating with these stakeholders may help AOM navigate the intersection of technology and healthcare, especially in a rapidly changing and highly regulated industry. Therefore, we rate AOM with a Level 4 Proactive Maturity for the GARP Principle of Compliance.

## 5.6 Principle of Availability

At AOM, Angela and James have his health data at their fingertips. From each step of the care process, they have access to timely, accurate, and efficiently retrieved information. Organizations that build on top of AWS may also contribute to data availability, creating a strong network effect that keeps patients engaged in Amazon's loop of services.

However, those outside of Amazon's suite of services or at its fringes may encounter difficulty having their information available to them. Like many immigrants, Angela and James were not signed up as Amazon Prime members when they first arrived in the US. Although they were fortunate enough to have paper records from the Philippines, it was far from the up-to-date electronic health records AOM providers expected them to have. Furthermore, their paper records were not always in the same format and could not be shared easily across Amazon's suite of services. Often more than once, Angela had to explain her challenges to the CS team and providers, disrupting continuity-of-care.

Angela and James' scenario shows that Amazon's strengths are confined within its spheres of influence. It cannot index patients that exist outside of its network, and nor can it serve them well. It is still just a business, not a government or country that can mandate people to sign up for its services. Even more so, Prime members still need a mobile device, internet connection, and technology literacy to use telehealth services. Patients with these challenges may find their experience at AOM reduced to that of a normal brick-and-mortar clinic, or perhaps, at the extreme end, that of a non-functional one that does not have the information it needs to work properly. Ultimately, AOM's business model operates on assumptions of privilege which ultimately hurt it and its patients.

To sufficiently scale and succeed in the business of global healthcare, Amazon must advocate for local, national, and international human rights and healthcare policies that bridge these assumptions. As an international company, it may look to international policy for guidance. The United Nations, for example, identifies 30 universally protected human rights (United Nations, 2022). One corollary to Article 25, "the right to a standard of living adequate for health and well-being", in the context of AOM, is the need for universal patient identifiers, a smartphone, and internet access. Although Amazon's international scale makes it especially susceptible to global human rights issues, it also makes it powerful enough to address them. With these gaps in mind, AOM scores Level 2 In Development Maturity for the GARP Principle of Availability.

**5.7 Principle of Retention**

Beverly appreciates how she can connect with her provider consistently through telehealth. Each virtual visit comes with email and text reminders beforehand, along with securely hosted summaries with provider notes afterwards. Outside of visits, she can scroll through all her notes online without concerns that they will get damaged or lost. Regardless of where she moves, her data will be stored long-term on the AWS cloud. As long as she has internet access, she can retrieve her health record and continue getting her primary care needs.

Though cloud services are convenient, there are times when Beverly is not able to use them. Without a phone plan, she relies on Wifi from friends and nearby coffee shops. While offline, she was upset when she found out that she missed an urgent message from one of her providers. Because AOM relies on the AWS cloud to exchange data between its suite of services, Beverly sometimes feels locked out of them without internet access. She tried printing or taking photos of her records, but she did not have the convenience of digital interaction and the security benefits that come with the cloud. Even though her data is retained by AWS, one of the most robust cloud storage providers with strong compliance and retention plans, these strengths are null without consistent access to technology.

Once again, Amazon's business model relies on a foundation of privilege that should not be assumed or taken for granted. We recommend similar steps on human rights and healthcare advocacy as stated above. In this context, AOM scores a Level 3 Essential Maturity for the GARP Principle of Retention.

**5.8 Principle of Disposition**

When Beverly finds an error in her medical record, she sends a message to her provider to fix it. Her provider confirms the mistake and accepts the correction. The change is then appended to her record by the CS team. Beverly asks the CS team about the process, and finds that older versions of the record are temporarily saved on the cloud, but will eventually be destroyed according to laws and regulations. When she visits a new AOM clinic out-of-state, she is pleased to see that all her updates are maintained across the unified cloud. Whether on her phone or at the point-of-care, the disposition process covers all records across Amazon's suite of services.

The frontend experience that satisfies Beverly's disposition needs is dependent on information governance at AWS. Although its commitments involve data controls and residency, data privacy, data sovereignty, and data security, there is currently no mention of data disposition (Data Protection, 2022). As a leading cloud service, AWS likely performs excellent disposition practices and protocols. However, even with strong backend infrastructure that may assist with disposition across services, disposition needs should be explicitly defined, consistent, and regularly evaluated (How the Information Governance Reference Model (IGRM) Complements ARMA International's Generally Accepted Recordkeeping Principles (GARP®), 2011).

Clarifying the disposition process to stakeholders will likely build confidence in this final step of the information lifecycle. As a result, we rate AOM with a Level 4 Proactive Maturity for the GARP Principle of Disposition.

**6.1 Conclusion**

Conventionally, IG is a framework to manage the multifaceted aspects of an organization's data and information. However, it is also a guide to understanding how the present may mesh with the future, such as long-term secular trends like telehealth. Ellis' story demonstrates the disruption of the traditional doctor-patient relationship, the rising role of ancillary service teams, the commoditization of healthcare, and the threats of monopolization and anti-competitive behavior. Meanwhile, Angela and James' scenario discusses the newfound security risks of convenience and patient empowerment that many take for granted. Protective safeguards, compliance, and regulation with stakeholders will be more important than ever. Navigating AOM also challenges assumptions of privilege and power that are often overlooked, exposing fundamental issues in human rights and healthcare policies that hurt all involved. Finally, Beverly's journey tests the end of the information lifecycle. She corroborates how technology may leave the underprivileged behind, and the urgent need to responsibly manage data, from creation to disposition. As the intersection between healthcare and technology grows, so, too, does the need for information management. Using IG, we may systematically analyze and synthesize each step and prepare organizations for the future.

**Addendum**

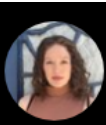

Exhibit 1. A Twitter meme of a hypothetical Amazon marketplace where users shop for healthcare providers (Mortensen, 2022).

| Name | Ellis | 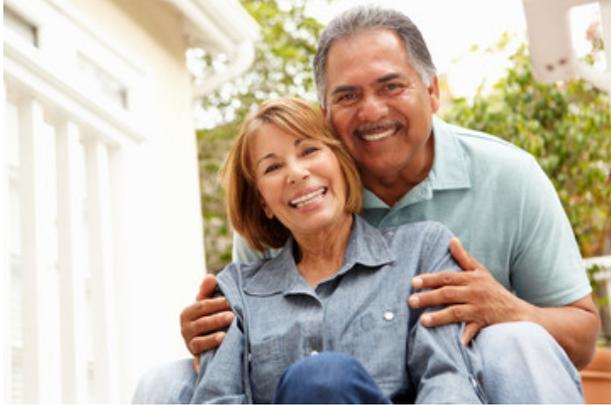 |
|---|---|---|
| Age | 70 | |
| Sex/Gender | Male | |
| Race/Ethnicity | Hispanic | |
| Marital Status | Married | |
| Education Level | High School | |
| Location | Austin, Texas | |
| Brief Description | Ellis, a military veteran who is now a retired carpenter, lives with his wife in Austin. He has hypertension and diabetes. In his spare time, he coaches a soccer team for kids. | |
| Trigger | During practice, Ellis suffers a heart attack. He promises his wife that he will be more accountable for his medications and diet regimen. He looks to AOM for a convenient, virtual solution to stay healthy. | |
| Scenario | As an Amazon Prime member, Ellis searches for primary care providers in his area on Amazon.com. The marketplace shows the name, location, price range, ratings, and reviews for each provider. He finds that picking and choosing the highest rated and most affordable provider is much easier than sticking with the same one from his last visit. After checking | |

| | out, he visits one of AOM's brick-and-mortar primary care offices for a blood test. He stays updated on prescriptions, diets, and exercises over telehealth. |
|---|---|
| Quote | "It's so easy to switch providers. I feel like I can shop for who best fits my needs on a trusted marketplace, but I do wish I had a better personal relationship with my providers." |

Table 1. Persona 1

| Name | Angela and her son, James | 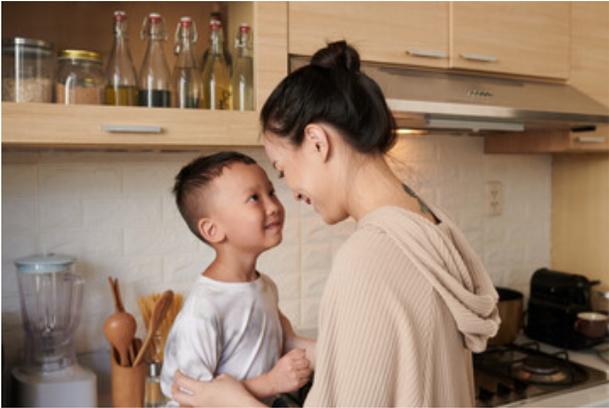 |
|---|---|---|
| Age | 30 and 6, respectively | |
| Sex/Gender | Female and Male, respectively | |
| Race/Ethnicity | Filipino-American | |
| Marital Status | Both are single | |
| Education Level | College | |
| Location | Los Angeles, CA | |

| Brief Description | Angela, a single mother, and her six-year-old son, James, are Filipino-American immigrants. To support the family, Angela is a work-from-home virtual assistant. At night, she takes classes to become a nurse. |
|---|---|
| Trigger | One day, James's school asks for his vaccine and immunization records. However, most of his medical records are back in the Philippines—and they're paper, not electronic. To accommodate her busy schedule, Angela navigates AOM to arrange vaccine and immunization services for her son. |
| Scenario | Recalling how she booked a health appointment on AOM for a client, Angela looks to the digital marketplace to get son's vaccine and immunizations updated for the coming school year. She enjoys having her son's documents on the cloud, but finds it difficult to retrieve her information from outside the US. However, she becomes concerned about AOM's convenience when realizes how many parties have access to her data across Amazon's suite of services. |
| Quote | "As a busy single-mother and immigrant, I want privacy, security, and safety for my family and me." |

Table 2. Persona 2

| Name | Beverly | 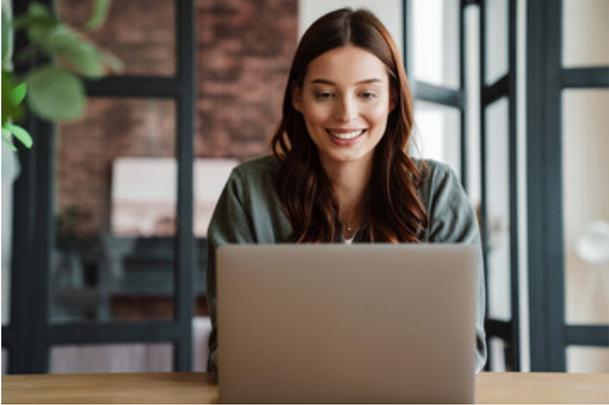 |
|---|---|---|
| Age | 28 | |
| Sex/Gender | Female | |
| Race/Ethnicity | White | |
| Marital Status | Single | |
| Education Level | College | |
| Location | San Francisco, CA | |
| Brief Description | Beverly was recently let go from her job in tech in San Francisco, CA. Her parents are unable to support her, so she couch surfs between friends nearby while searching for employment. The stress aggravates her asthma. | |
| Trigger | Without insurance, she looks to AOM for affordable primary care to work on her respiratory health. She hopes that the marketplace can match her with an empathetic provider that has time to listen to her needs. | |

| | |
|---|---|
| Scenario | Beverly is fortunate to have Wifi at coffee shops and her friend's apartment, but she worries that she may not have consistent internet for telehealth like she expected. Time-sensitive messages with providers are often sent or received late, which makes it difficult to coordinate care. When she finds out there is incorrect information on her record, she contacts customer service to alter it. |
| Quote | "Technology is great, but only if I am able to use it and have access to it in the first place." |

Table 3. Persona 3